\documentclass[pdftex,twocolumn,epjc3]{svjour3}          

\RequirePackage[T1]{fontenc}

\smartqed  

\RequirePackage{graphicx}
\RequirePackage{mathptmx}      
\RequirePackage{flushend}
\RequirePackage{mathrsfs} 
\RequirePackage[overload]{textcase} 
\RequirePackage{amssymb} 
\RequirePackage{subfigure}
\RequirePackage{gensymb}
\RequirePackage{upgreek}
\RequirePackage{dcolumn}
\RequirePackage{bm}
\RequirePackage{nth}
\RequirePackage[numbers,sort&compress]{natbib}
\RequirePackage[colorlinks,citecolor=blue,urlcolor=blue,linkcolor=blue]{hyperref}

\begin{document}

\title{Discovery potential for directional Dark Matter detection with nuclear emulsions}

\subtitle{NEWSdm Collaboration}

\author{N.~Agafonova\thanksref{INR}
        \and
        A.~Aleksandrov\thanksref{INFNNA,UNINA}
         \and
         A.~Anokhina\thanksref{SINP}
         \and
         T.~Asada\thanksref{NAGOYA}
         \and
         V.V.~Ashikhmin\thanksref{INR}
         \and
         I.~Bodnarchuk\thanksref{JINR}
         \and
         A.~Buonaura\thanksref{INFNNA,UNINA}
          \and
         M.~Chernyavskii\thanksref{LPI}
         \and
         A.~Chukanov\thanksref{JINR}
         \and
         N.~D'Ambrosio\thanksref{LNGS}
         \and
         G.~De Lellis\thanksref{INFNNA,UNINA}
         \and
         A.~Di Crescenzo\thanksref{t1,e1,INFNNA,UNINA}
         \and 
         N.~Di Marco\thanksref{t1,e2,LNGS}
         \and
         S.~Dmitrievski\thanksref{JINR}
         \and
         R.I.~Enikeev\thanksref{INR}
         \and
         R.~A.~Fini\thanksref{INFNBA}
         \and
         G.~Galati\thanksref{INFNNA,UNINA}
         \and
         V.~Gentile\thanksref{GSSI}
         \and 
         S.~Gorbunov\thanksref{LPI}
         \and
         Y.~Gornushkin\thanksref{JINR}
         \and
         A.~M.~Guler\thanksref{METU}
         \and
         H.~Ichiki\thanksref{NAGOYA}
         \and
         T.~Katsuragawa\thanksref{NAGOYA}
         \and
         N.~Konovalova\thanksref{LPI}
	\and
	K.~Kuge\thanksref{CHIBA}
	\and
	A.~Lauria\thanksref{INFNNA,UNINA}
	\and
	K.Y.~Lee\thanksref{KOREA}
	\and
	L.~Lista\thanksref{INFNNA}
	\and
	A.S.~Malgin\thanksref{INR}
	\and
	A.~Managadze\thanksref{SINP}
	\and
	P.~Monacelli\thanksref{INFNRO}
	\and
	M.~C.~Montesi\thanksref{INFNNA,UNINA}
	\and
	T.~Naka\thanksref{NAGOYA}
	\and
	N.~Okateva\thanksref{LPI}
	\and
	B.D.~Park\thanksref{KOREA}
	\and
	D.~Podgrudkov\thanksref{SINP}
	\and
	N.~Polukhina\thanksref{LPI,NRNU}
	\and
	F.~Pupilli\thanksref{INFNPD,UNIPD}
	\and
	T.~Roganova\thanksref{SINP}
	\and
	A.~Rogozhnikov\thanksref{YANDEX}
	\and
	G.~Rosa\thanksref{INFNRO,UNIRO}
	\and
	O.G.~Ryazhskaya\thanksref{INR}
	\and
	O.~Sato\thanksref{NAGOYA}
	\and
	I.R.~Shakiryanova\thanksref{INR}
	\and
	T.~Shchedrina\thanksref{LPI}
	\and
	C.~Sirignano\thanksref{INFNPD,UNIPD}
	\and
	J.Y.~Sohn\thanksref{KOREA}
	\and
	A.~Sotnikov\thanksref{JINR}
	\and
	N.~Starkov\thanksref{LPI}
	\and
	P.~Strolin\thanksref{INFNNA,UNINA}
	\and
	V.~Tioukov\thanksref{INFNNA,UNINA}
	\and
	A.~Umemoto\thanksref{NAGOYA}
	\and
	A.~Ustyuzhanin\thanksref{YANDEX,NRU}
	\and
	C.S.~Yoon\thanksref{KOREA}
	\and
	M.~Yoshimoto\thanksref{NAGOYA}
	\and
	S.~Vasina\thanksref{JINR}
}

\thankstext[$\star$]{t1}{Corresponding author}
\thankstext{e1}{e-mail: antonia.dicrescenzo@na.infn.it}
\thankstext{e2}{e-mail: natalia.dimarco@lngs.infn.it}

\institute{INR RAS-Institute for Nuclear Research of the Russian Academy of Sciences, Moscow, Russia\label{INR}
          \and
          INFN Sezione di Napoli, Napoli, Italy\label{INFNNA}
          \and
         Dipartimento di Fisica dell'Universit\`a Federico II di  Napoli, Napoli, Italy \label{UNINA}
         \and
         INP MSU-Skobeltsyn Institute of Nuclear Physics of Moscow State University, Russia \label{SINP}
         \and
         Nagoya University and KM Institute, Nagoya, Japan\label{NAGOYA}
         \and
         JINR-Joint Institute for Nuclear Research, Dubna, Russia\label{JINR}
         \and
         LPI-Lebedev Physical Institute of the Russian Academy of Sciences, Moscow, Russia\label{LPI}
         \and
         INFN-Laboratori Nazionali del Gran Sasso, Assergi (L'Aquila), Italy \label{LNGS}
         \and
         INFN Sezione di Bari, Bari, Italy\label{INFNBA}
         \and
         Gran Sasso Science Institute, L'Aquila, Italy\label{GSSI}
         \and
         METU-Middle East Technical University, Ankara, Turkey\label{METU}
         \and
         Chiba University, Chiba, Japan\label{CHIBA}
         \and
         RINS and Department of Physics Education, Gyeongsang National University, Jinju, Korea \label{KOREA}
         \and
         INFN Sezione di Roma, Roma, Italy\label{INFNRO}
         \and
         National Research Nuclear University - Moscow Engineering Physical University, Moscow, Russia\label{NRNU}
         \and
	INFN Sezione di Padova, Padova, Italy\label{INFNPD}
	\and
	Dipartimento di Fisica e Astronomia dell'Universit\`a di Padova, Padova, Italy\label{UNIPD}
	\and
	Yandex School of Data Analysis, Moscow, Russia\label{YANDEX}
	\and
	Dipartimento di Fisica dell'Universit\`a di Roma, Rome, Italy\label{UNIRO}
	\and
	National Research University, Higher School of Economics, Moscow, Russia\label{NRU}
}


\maketitle

\begin{abstract}
Direct Dark Matter searches are nowadays one of the most fervid research topics with many experimental efforts devoted to the search for nuclear recoils induced by the scattering of Weakly Interactive Massive Particles (WIMPs). 
Detectors able to reconstruct the direction of the nucleus recoiling against the scattering WIMP are opening a new frontier to possibly extend Dark Matter searches beyond the neutrino background. Exploiting directionality would also prove the galactic origin of Dark Matter with an unambiguous signal-to-background separation. Indeed, the angular distribution of recoiled nuclei is centered around the direction of the Cygnus constellation, while the background distribution is expected to be isotropic. 
Current directional experiments are based on gas TPC whose sensitivity is  limited by the small achievable detector mass. In this paper we present the discovery potential of a directional experiment based on the use of a solid target made of newly developed nuclear emulsions and of optical read-out systems reaching unprecedented nanometric resolution.
\end{abstract}

\section{\label{sec:intro} Introduction}

Although its nature remains undisclosed, there are indications that the Dark Matter (DM) is made by hypothetical particles called WIMPs (Weakly Interacting Massive Particles). 
Current experimental efforts in the field of direct DM searches are devoted to the detection 
of the rare interactions of WIMPs
from the galactic halo with nuclei in a terrestrial detector.

As DM detectors are rapidly improving in sensitivity, at some point they will encounter the so-called ``neutrino floor'' 
where coherent scattering of Solar, atmospheric, and diffuse supernovae neutrinos creates an irreducible background. Neutrinos are therefore the ultimate background for WIMP direct detection searches as they produce recoils with similar rates and energy spectra.

New generation detectors capable of measuring the direction of nuclear recoils induced by
WIMP elastic scatterings would provide an unambiguous identification of WIMPs as candidate
for the galactic DM and could overcome the limit imposed by coherent neutrino scattering.

Several directional approaches have been proposed \cite{Battat}.  Experiments based on the use of low pressure gaseous Time Projection Chambers (TPCs) typically make use of compounds rich in $^{19}$F nucleus in order to set limits on the Spin-Dependent (SD) coupling of WIMPs with less than kg mass
detectors \cite{Tovey}.
Nevertheless this technology is hardly scalable to very large detector masses needed to reach a good sensitivity to the Spin-Independent (SI) case. 

The use of a solid target for directional searches would  overcome the mass limitation of the gaseous TPC approach thus allowing to reach a high sensitivity in the low cross section sectors of the SI case. Nevertheless, in a solid medium, the track of the WIMP-scattered nuclear recoil will have a path length of the order of a few hundred nanometers. A detector with unprecedented high tracking resolution is therefore needed: the approach proposed  by the NEWSdm  Collaboration consists of using  a nuclear emulsion-based detector acting both as target and as tracking device. The project foresees the use of a novel emulsion technology called Nano Imaging Tracker (NIT) \cite{Natsume,Naka} featuring a position resolution an order of magnitude higher than that of the emulsion used in the OPERA experiment ($\sim$1$\mu$m) \cite{Acquafredda}. The detector is made of a stack of NIT emulsion films surrounded by a shield to reduce the external background. The detector is then placed on an equatorial telescope in order to keep fixed the detector orientation with respect to the incoming apparent WIMP flux, i.e.~toward the Cygnus constellation, thus compensating the Earth rotation. The emulsion films are placed in such a way that their surface is permanently parallel to the Galactic plane.
The angular distribution of the WIMP-scattered nuclei is therefore expected to be strongly anisotropic with a peak centered in the opposite direction to the motion of the Solar System in the Galaxy.  A detailed description of the NEWSdm experiment can be found in Ref.~\cite{Aleksandrov}. The capability to detect nuclear recoils like those induced by a WIMP
scattering was demonstrated with NIT emulsion films and an intrinsic
angular resolution of 13 degrees was measured \cite{Aleksandrov}.

In this paper we describe the potentialities of an emulsion-based detector as foreseen by the NEWSdm experiment, that is capable of distinguishing the anisotropic WIMP signal over the isotropic background. We report the discovery potential and the exclusion limit for a typical detector mass per exposure time of 100 kg$\cdot$year and we give future prospects towards the overcome of the neutrino background limit.

\section{\label{sec:analysis} Signal and background modelling}
\subsection{\label{sec:model} Dark Matter model}
We base our analysis on the hypothesis of a Standard Halo Model (SHM) describing the form of the DM distribution. For a spherically symmetric-isothermal DM halo, with density profile $\rho(r) \propto r^{-2}$, the resulting velocity distribution in the reference frame of the Galaxy  has a Maxwellian--Boltzmann form \cite{Ullio}
\begin{equation}
f({\bf v}) = \frac{1}{(2\pi \sigma_v^2)^{1/2}}\exp \biggl( -\frac{ {\bf v}^2}{2\sigma_v^2}\biggr)
\end{equation}
with velocity dispersion $\sigma_v = v_0/\sqrt{2}$ and $v_0\simeq220$ km$\cdot$s$^{-1}$. A cut-off in the distribution at the Galactic escape velocity $v_{esc}\simeq544$ km$\cdot$s$^{-1}$ is considered. 

We take a detector velocity equal to the tangential component of the Sun motion around the Galactic center $v_{Earth}=220$ km$\cdot$s$^{-1}$, and we neglect the Sun peculiar velocity with respect to the
Local Standard of Rest (LSR) and the Earth orbital velocity around the Sun, their contribution being smaller than the uncertainty on the Sun velocity \cite{Morgan}.

\subsection{\label{sec:signal}Signal and background detection}

In this work we consider the realistic NEWSdm detector with a 100 nm threshold and a 2D track reconstruction capability, as proven by test beam results reported in \cite{Aleksandrov}. 
Studies are ongoing to further reduce the threshold, as outline in the last section of this work. 
Additional improvements devoted to exploit the intrinsic emulsion capability of recording 3D tracks with nanometric accuracy are the subject of a future experimental program and are not taken into account in the present work.  

\begin{figure}[hbtp]
\centering
\begin{center}
\includegraphics[width=9cm]
{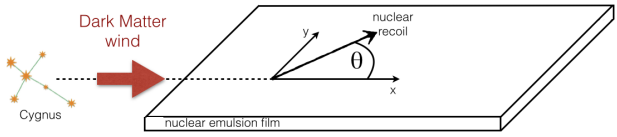}
\end{center}
\caption{\label{fig:system} The 2D reconstruction is performed in the $xy$-plane. The $x$-axis is directed opposite to the Cygnus constellation and $\theta$ is the angle between the $x$-axis and the projection of the nuclear recoil in the $xy$-plane.}
\end{figure}

The track reconstruction is performed in the $xy$ Galactic plane, where the $x$-axis is directed opposite to the Cygnus constellation (see Figure \ref{fig:system}). In this reference system, the angle of the nuclear recoil ($\theta$) is defined as the  angular difference between its projection in $xy$-plane and the $x$-axis. Assuming no head-tail sense recognition, the 2D angle is lies in the $[-\pi/2,\pi/2]$ range.

The angular distribution of the recoiled nuclei is expected to have a Gaussian shape peaked at zero, with standard deviation  ($\sigma_\theta^{Wn-scatt}$) that decreases as the WIMP mass increases. The lighter the WIMP, the stronger the angular aniso\-tropy. 
Indeed, due to the fact that low WIMP mass induce an energy distribution shifted to low energy, events above 100 nm threshold are those with the most pronounced directional feature. 

As far as the background is concerned, the contribution from neutrons is usually considered as irreducible, since they induce nuclear recoils as WIMPs do. Neutrons in underground laboratories originate  from cosmic muon interactions, environmental radioactivity,  spontaneous fissions and ($\alpha$, n) reactions \cite{Aleksandrov2}. The first two sources can be reduced by an appropriate shielding; the latter one, coming from the intrinsic radioactivity of the target materials, is associated with an isotropic distribution in the laboratory frame.  

Background from $\beta$-rays produced in $^{14}$C decay is expected to be negligible thanks to the different response of emulsions to electrons and nuclear recoils and to the use of gelatine with low C content or of synthetic polymers instead. Any eventual residual background component from $\beta$-rays  would anyway be isotropic.

The directional detection has the unique capability of distinguishing the WIMP signal from the background by exploiting the feature of the signal, expected to be peaked in the opposite direction to the motion of the Sun. 

\subsection{\label{sec:recoil}Recoil simulation}

Nuclear recoil tracks produced in WIMP interactions show deviations from their original path due to continuous collisions with atoms in the target. In order to quantify these deviations we evaluated the effect of the straggling in an emulsion target made of the following nuclei (mass fraction
in \%): H (1.6), C (10.1), N (2.7), O (7.4), S (0.3), Br (32.0), Ag (44.0), I (1.9). The total density is 3.43 g/cm$^3$.
When a charged particle passes through the emulsion film, a number of AgBr grains per unit path length are sensitised.  Measurements performed with both a SEM and an X-ray microscope showed that protons or heavier nuclei produce on average 12 grains/$\mu$m in NIT emulsions. 

The study was performed with the SRIM \cite{Ziegler} software package and its TRIM (Transport of Ions in Matter) track generator Monte Carlo program. 

All nuclear recoils were simulated in the energy range from 0 to 500 keV. All the tracks with at least 100 nm length were considered.  The range was calculated using the coordinates of the first and last point of the ion track. 
The energy threshold corresponding to the minimum track length was evaluated from the average range estimated by SRIM. It ranges from 25 keV for Carbon to 273 KeV for Iodine.
Being NIT emulsions less sensitive to protons, H recoils are conservatively not taken into account.

\begin{figure}[hbtp]
\centering
\begin{center}
\includegraphics[width=7cm]
{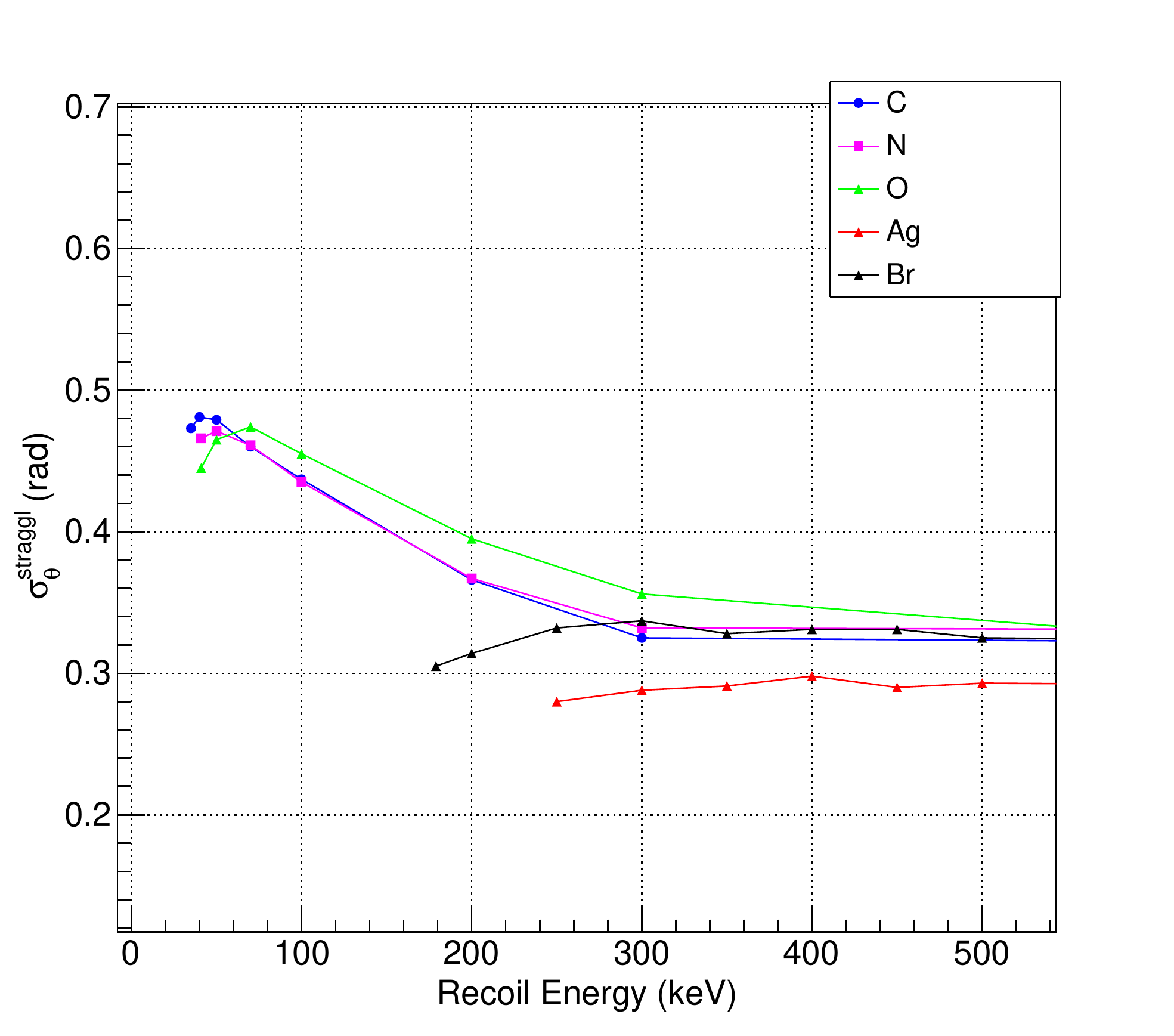}
\end{center}
\caption{\label{fig:scattering} Angular deviation due to the straggling of nuclei in an emulsion target. A 100 nm threshold on the  range of the recoil track is applied.}
\end{figure}

In the above mentioned DM model, the mean direction of the track was compared with the initial direction of the recoil: the difference of the two directions shows a Gaussian distribution centered at zero, with a sigma ($\sigma_\theta^{straggl}$) that depends on the selected nuclei and on the recoil energy and reflects the loss of the directional information due to the straggling in the target. Figure \ref{fig:scattering} shows $\sigma_\theta^{straggl}$ as a function of the recoil energy for the most abundant nuclei in emulsion. Only ranges above 100 nm are considered.
The angular deviation due to straggling ranges from 0.3 to 0.5 rad for light nuclei (C, N, O) and from 0.2 to 0.35 for heavy nuclei (Ag, Br).

\begin{figure}[hbtp]
\centering
\begin{center}
\includegraphics[width=7cm]
{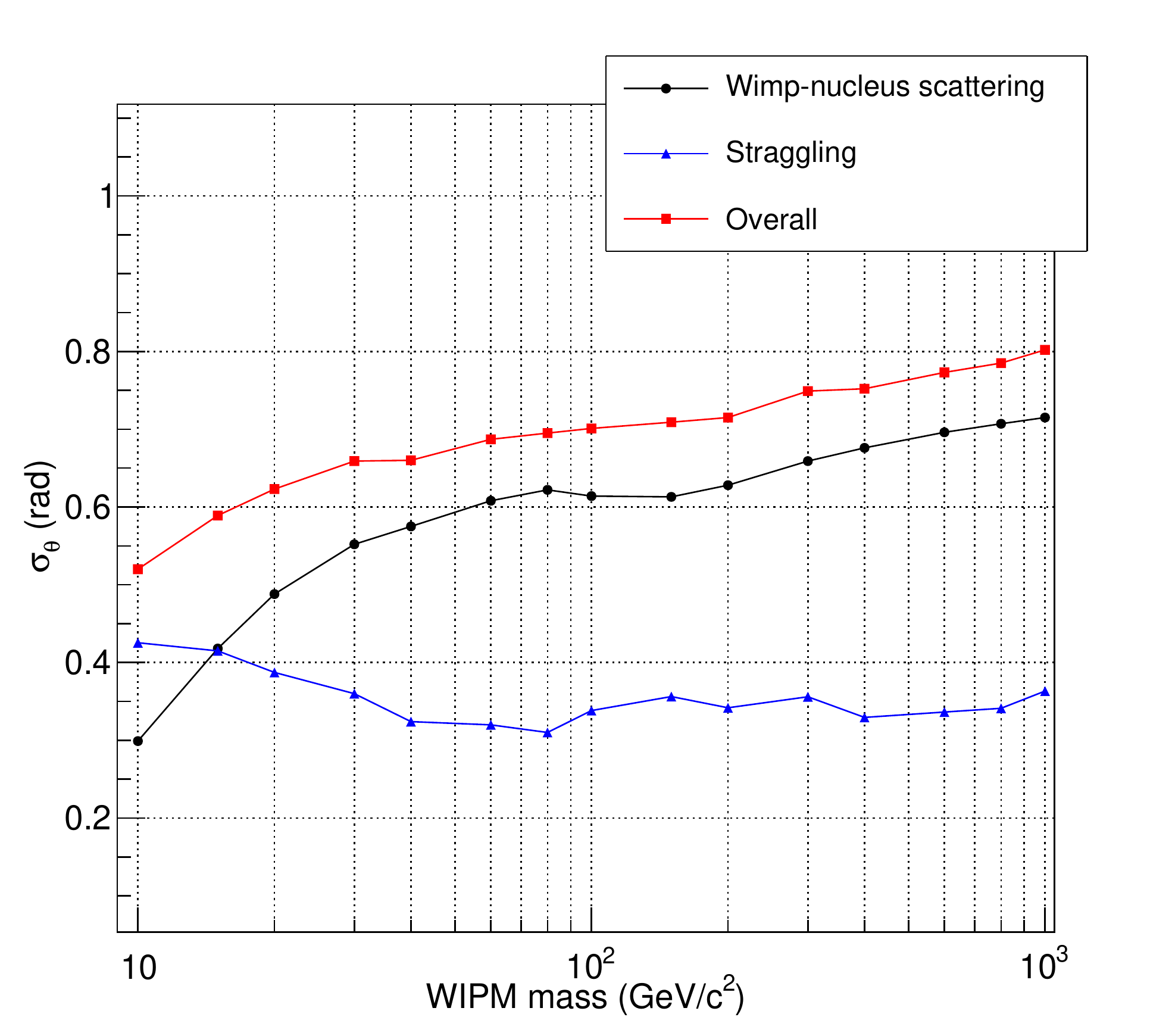}
\end{center}
\caption{\label{fig:resolution} Overall angular deviation of nuclear recoils in an emulsion target as a function of the WIMP mass (red boxes). The two components due to the WIM-nucleus scattering (black dots) and the straggling (blue triangles) are also shown.}
\end{figure}

Figure \ref{fig:resolution} shows the overall angular deviation expected for nuclear recoils in an emulsion target as a function of the WIMP mass. Black dots represent the angular deviation of nuclear recoils from the direction of Earth's motion, as it arises from the WIMP-nucleus scattering ($\sigma_\theta^{Wn-scatt}$), the blue triangles represent the angular deviation due to straggling in emulsion target only ($\sigma_\theta^{straggl}$), the red boxes are the convolution of the two components, hereafter referred to as $\sigma_\theta^{TOT}$
\begin{displaymath}
\sigma_\theta^{TOT} = \sqrt{\left(\sigma_\theta^{Wn-scatt}\right)^2+\left(\sigma_\theta^{straggl}\right)^2} .
\end{displaymath}

\section{\label{sec:method}Signal extraction}

We used the variable  $\sigma_\theta^{TOT}$ as a function of the incident WIMP mass to evaluate the capability of an emulsion directional detector to distinguish the WIMP-induced signal from the expected isotropically distributed background.

\begin{figure}[hbtp]
\centering
\begin{center}
\includegraphics[width=7cm]
{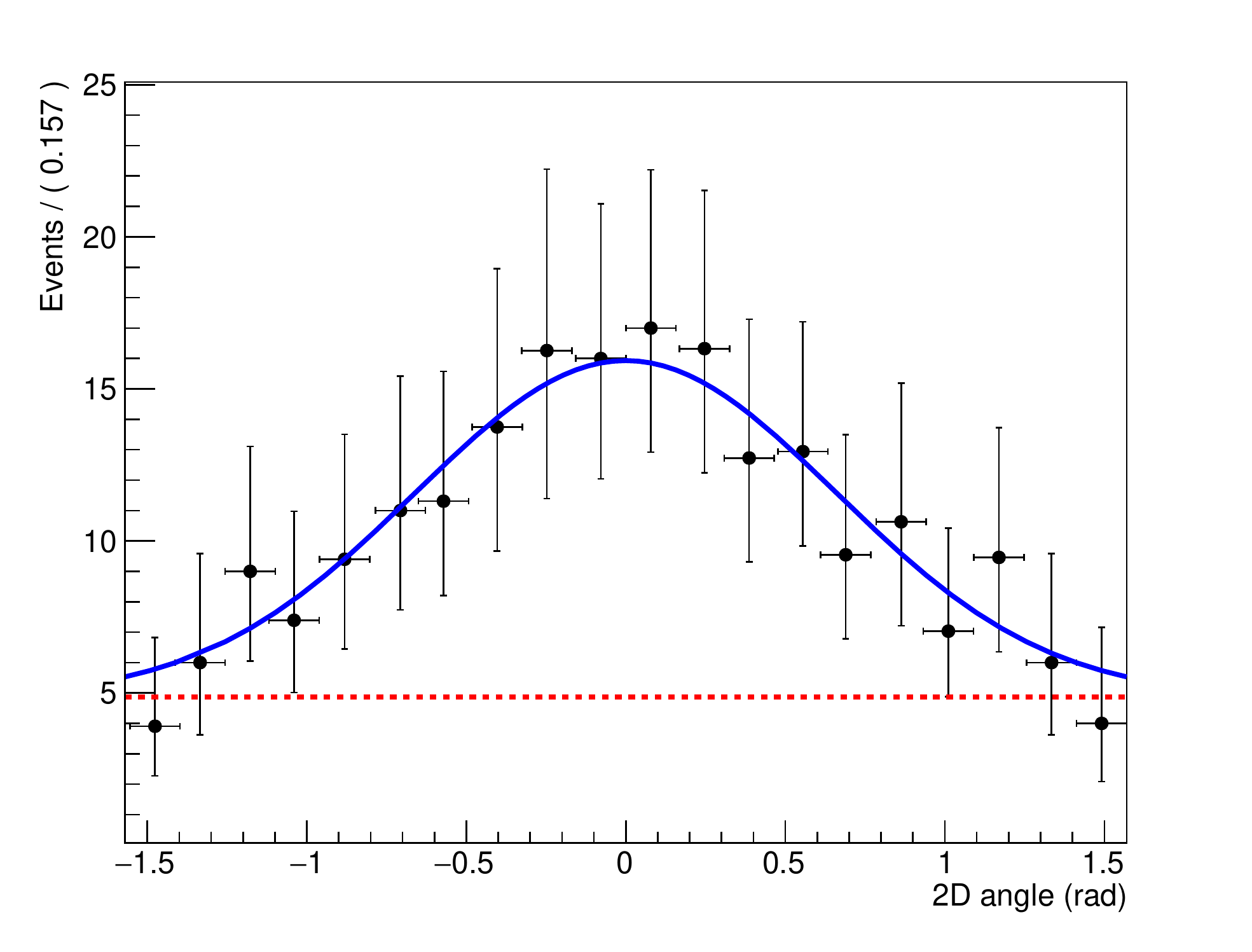}
\end{center}
\caption{\label{fig:model} 2D angular distribution of 100 WIMP-induced recoils and 100 background events. Signal and background components are represented as solid blue and dashed red lines, respectively. Recoils are produced for an emulsion target, a 40 GeV/c$^2$ mass WIMP and a 100 nm threshold.}
\end{figure}

The estimate of the expected significance of a directional emulsion experiment was performed in a frequentist approach with the Profile Likelihood ratio test, taking into account the characterisation of the signal angular spectrum, as a function of the incident WIMP mass. In the Profile Likelihood ratio test \cite{Cowan} the null hypothesis $H_{0}$ (background only) is tested against the alternative hypothesis $H_{1}$ including both signal and background. 
We have considered the extended likelihood function 
\begin{equation}
 \mathscr{L} = \frac{(\mu_s+\mu_b)^N}{N!} e^{-(\mu_s+\mu_b)} \times
\end{equation}
\begin{displaymath}
\times \prod_{i=1}^{N} \left( \frac{\mu_s}{\mu_s+\mu_b} S(R_i) + \frac{\mu_b}{\mu_s+\mu_b}B(R_i) \right)
\end{displaymath}
where $\mu_b$ and $\mu_s$  are the number of expected background and WIMP events, respectively; $N$ is the total number of observed events,  $R_i$ is the direction of each event while the functions $S$ and $B$ are the probability density functions (PDF) for signal and background, respectively. Figure \ref{fig:model} shows an example of the model used to describe signal and background: the signal PDF is a Gaussian distribution of the 2D angular recoils  for a WIMP mass of 40 GeV/c$^2$ (blue curve) while the background PDF is a uniform distribution (dashed red line). Both distributions are normalized to 100 events.

The statistical model was implemented within the RooStats tool \cite{Moneta}, whose classes are built on top of the RooFit \cite{Verkerke} package  of the ROOT framework \cite{Brun}.

\section{\label{sec:discovery}Discovery potential}

In the following we consider a realistic 100 kg$\cdot$year NEWSdm detector with a recoil track length range above 100 nm and no head-tail sense recognition.

\begin{figure}[hbtp]
\centering
\begin{center}
\includegraphics[width=7cm]
{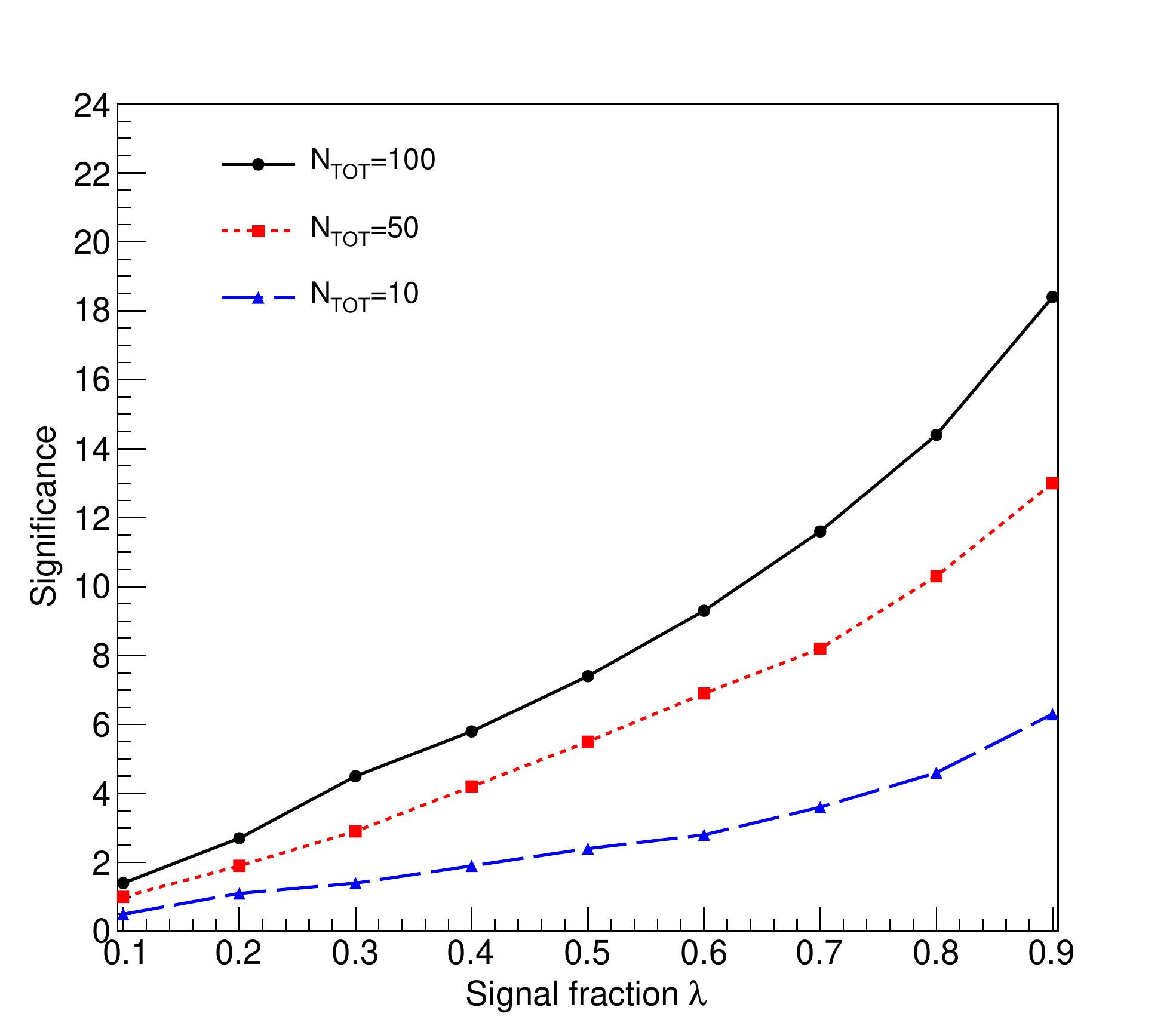}
\end{center}
\caption{\label{fig:sig_lambda} Mean significance as a function of the expected WIMP fraction $\lambda$ for $N$ = 10 (black solid line), 50 (red dotted line), 100  (blue dashed line). A 40 GeV/c$^2$ WIMP mass is assumed. 
}
\end{figure}

\begin{figure}[hbtp]
\centering
\begin{center}
\includegraphics[width=7cm]
{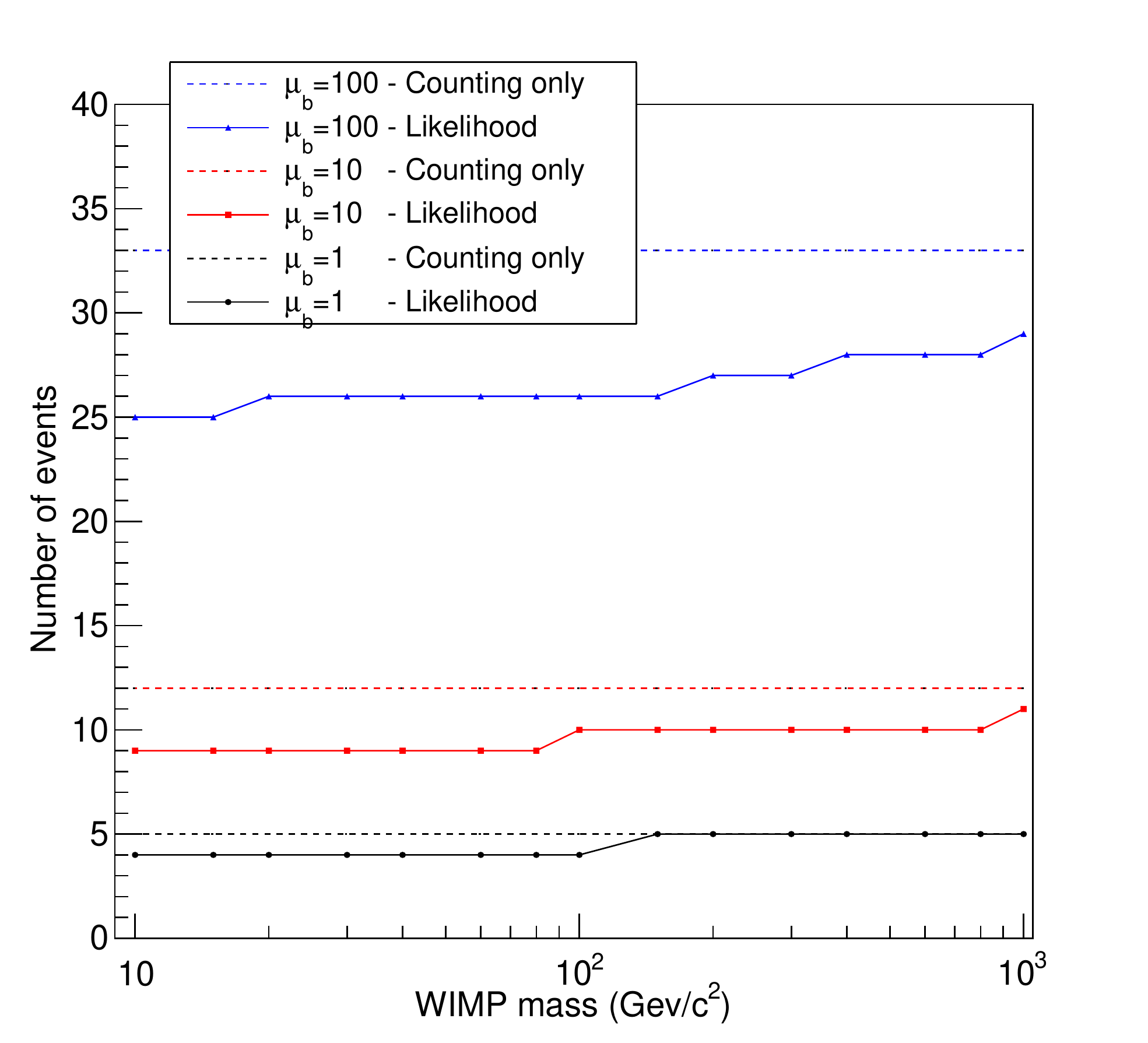}
\end{center}
\caption{\label{fig:nevents} Minimum number of signal events required to get a 3$\sigma$ evidence as a function of the WIMP mass. Three background hypotheses are considered: 100 (black line), 10 (red line) and 1 (blue line) events. Dotted lines show the results obtained with the counting method.}
\end{figure}

We studied the effect of background/signal contribution to data. Figure \ref{fig:sig_lambda} shows the behavior of the mean significance as a function of the data purity $\lambda$  = $\mu_s/(\mu_s+\mu_b)$, i.e.~the expected fraction of signal events in the data, for three different  numbers of observed events $N = 10, 50, 100$, for a WIMP mass of 40 GeV/c$^2$.  As expected the mean significance is an increasing function of $\lambda$. In addition, for a fixed $\lambda$, increasing the total number of events determines an improvement in the significance. For a given value of $\lambda=0.5$, for example, the significance of a DM detection can be enhanced from $2\sigma$ to $6\sigma$ with an exposure ten times larger.
Figure \ref{fig:nevents} reports the minimum number of signal events required to achieve a 3$\sigma$ evidence as a function of the WIMP mass. Three cases are presented: 1, 10, 100  background events. For comparison, the curves obtained with a counting method based on the Poisson statistics are also reported. The counting method is the most conservative approach for the significance evaluation because it does not assume any knowledge on the shape of both background and signal.
In case of a large background contamination ($\mu_b$=100), at high WIMP masses ($\sim$1000 GeV/c$^2$) about 29 signal events are required to claim a 3$\sigma$ evidence while about 25 signal events are enough at low WIMP masses ($\sim$6 GeV/c$^2$).

The required number of signal events decreases as the background contamination is reduced, ranging from 11 to 9 events for $\mu_b$=10 and from 5 to 4 events for $\mu_b$=1.

\begin{figure}[hbtp]
\centering
\begin{center}
\includegraphics[width=7cm]
{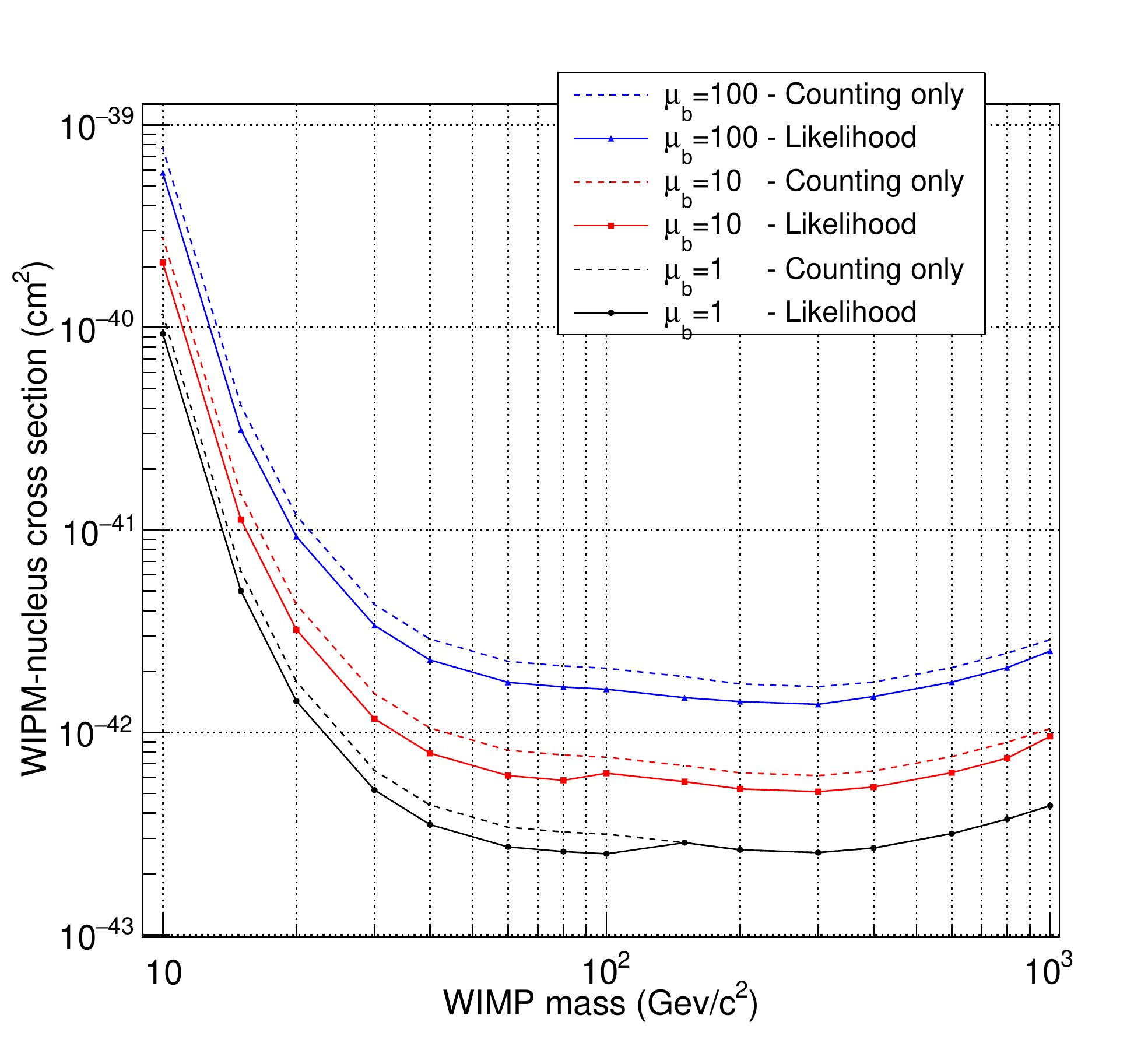}
\end{center}
\caption{\label{fig:sig_all} 3$\sigma$ evidence curve in the (mass, cross section) plane for a 100 kg$\cdot$year emulsion detector.  Three background hypotheses are considered: 100 (black line), 10 (red line) and 1 (blue line) events. Dotted lines show the results obtained with the counting method.}
\end{figure}

The 3$\sigma$  curve in the (mass, cross section)  plane was obtained for the three background hypotheses, assuming a 100 kg$\cdot$year exposure: the results obtained with the likelihood ratio test are reported in Figure \ref{fig:sig_all}, together with the ones obtained with the counting method. One can see that a directional detector may lead to an improvement with respect to the counting method ranging from 10\% to 20\% (8\% to 25\%) for $\mu_b$=100 ($\mu_b$=10).

\section{\label{sec:upperlimit}Upper limit evaluation}

In case no evidence for DM can be drawn from data, an exclusion limit can be derived. This will be the case for the very first results of an emulsion directional detector with short exposure.

\begin{figure}[hbtp]
\centering
\begin{center}
\includegraphics[width=7cm]
{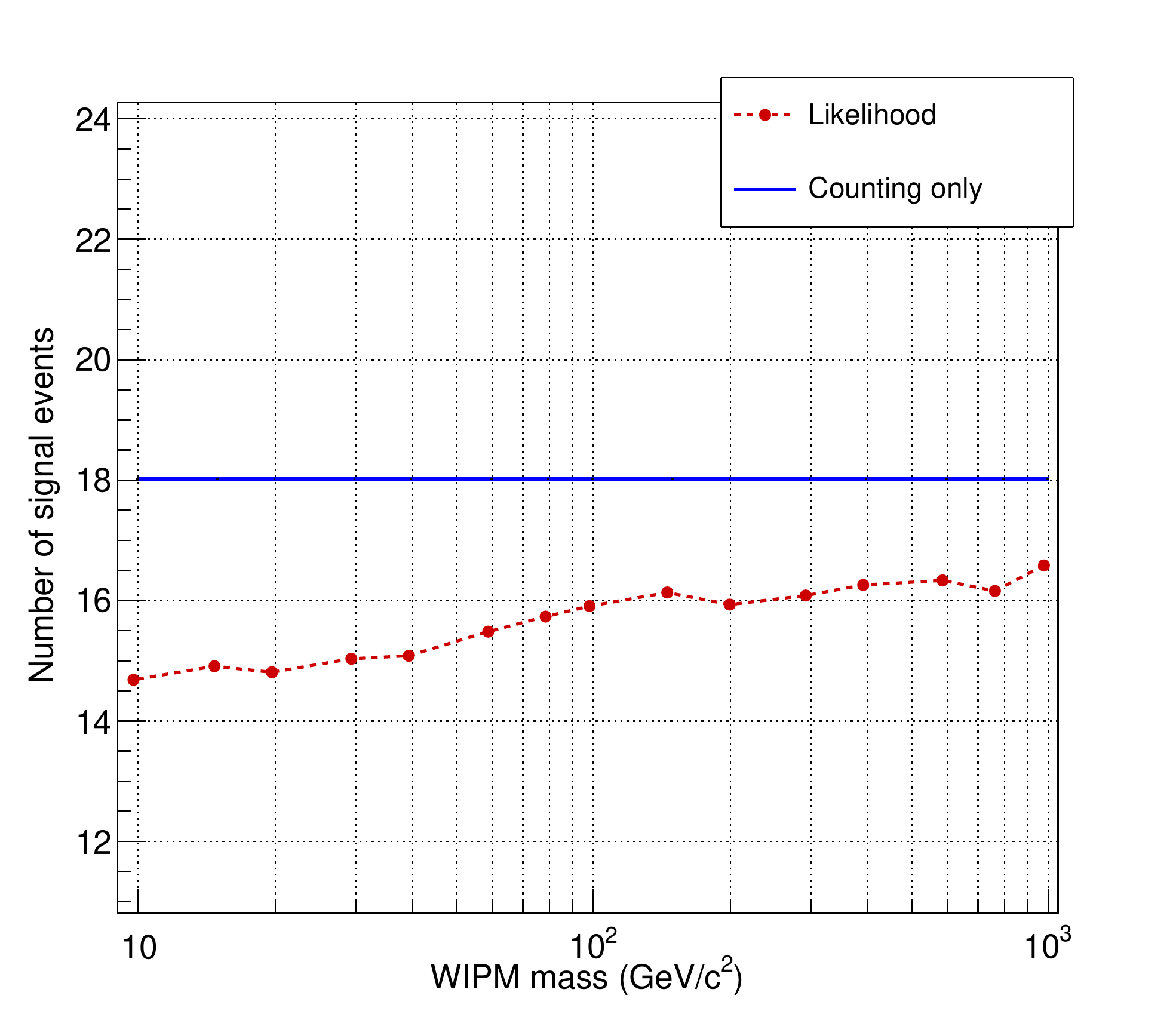}
\end{center}
\caption{\label{fig:Nevents_b100}  Upper limit on the number of signal events at 90\% CL for WIMP masses in the [10, 1000] GeV/c$^2$ range for $\mu_b=100$. The solid blue line represents the upper limit obtained with the counting method.}
\end{figure}

For the upper limit evaluation we have used the Profiled Likelihood Ratio method, with modified frequentist approach (CLs method \cite{Read}), and we have fixed the Confidence Level at 90\%. The sensitivity curve is achieved by finding the exclusion limits for the different WIMP masses. Figure \ref{fig:Nevents_b100} shows the results of the scan of WIMP masses in the range [10, 1000] GeV/c$^2$ performed with the Likelihood Ratio method. It represents the upper limit to the number of signal events for 100 expected background events, ranging from about 15 to 17 events. For comparison, the curve obtained with the counting method is also reported.

\begin{figure}[hbtp]
\centering
\begin{center}
\includegraphics[width=7cm]
{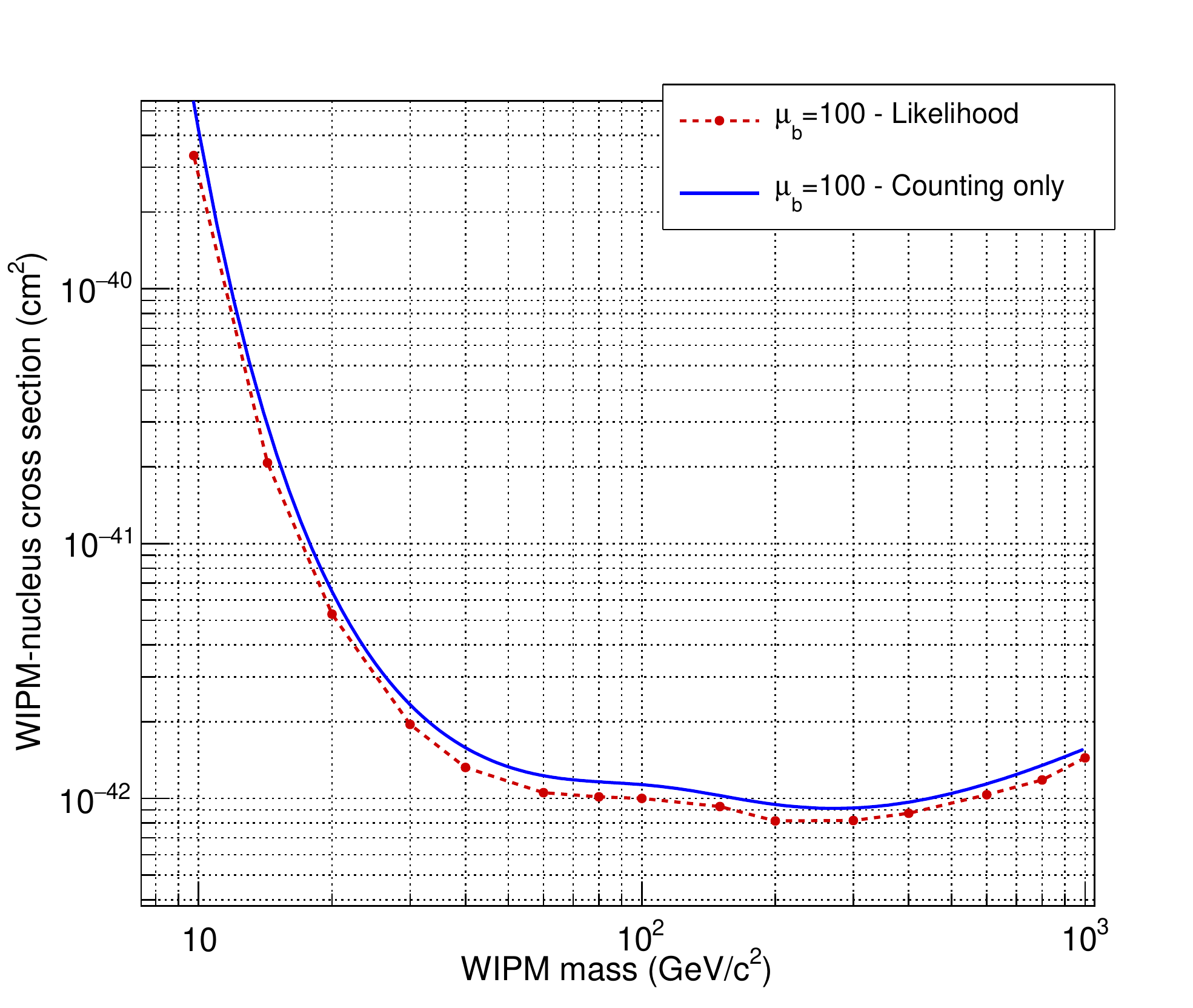}
\end{center}
\caption{\label{fig:excl_100}  Exclusion curve at 90\% CL for WIMP in the (mass, cross-section) plane for a 100 kg$\cdot$year emulsion detector and $\mu_b=100$. The continuous blue line represents the exclusion curve obtained with the counting method.}
\end{figure}

The upper limit on the number of signal events can be translated into the exclusion limit in the (mass, cross section) plane. Figure \ref{fig:excl_100} shows the exclusion curve obtained assuming a 100 kg$\cdot$year exposure with 100 background events. The improvement with respect to the counting method ranges from 10\% to 20\%.

\section{\label{sec:identification}Dark Matter identification}

Beyond the detection of an excess above the background expectation, the demonstration that the signal is due to Dark Matter stands as a major challenge.

A directional experiment has the unique opportunity to test the anisotropy of the observed signal, thus providing an unambiguous proof of the WIMP origin of the recoil signal. The measurement of the recoil direction is, indeed, a ``smoking gun'' for the rejection of the isotropic background hypothesis. 

We have used a likelihood ratio test to quantitatively estimate the separation between the signal and background hypotheses.
The likelihood functions in the background null hypothesis $H_0$ and in the signal  hypothesis $H_1$ are defined, respectively, as 
 \begin{equation}
 \mathscr{L}_b \equiv   \mathscr{L}(x_1, ... , x_n \mid H_0) =  \prod_{i=1}^{N} f(x_i\mid H_0)
\end{equation}
\begin{equation}
 \mathscr{L}_s \equiv   \mathscr{L}(x_1, ... , x_n \mid H_1) =  \prod_{i=1}^{N} f(x_i\mid H_1) 
 \end{equation}
where $N$ is the number of observed events, $x_i$ the measured recoil direction for the \(i\)-th  event, $f(x_i\mid H_0)$ and $f(x_i\mid H_1)$ the probability to observe the value $x_i$ under the background only and the WIMP only hypothesis, respectively.

We have used the signal characterisation as a function of the WIMP mass reported in Sec.~\ref{sec:method}, including the straggling effect of nuclear recoils in the emulsion target. 

For a given number of observed events $N$ we simulated 10$^4$ pseudo-experiments and evaluated the test statistics defined as 
\begin{equation}
q= 2 \mbox{ln}( \mathscr{L}_s/ \mathscr{L}_b) 
\end{equation}

\begin{figure}[hbtp]
\centering
\begin{center}
\includegraphics[width=7cm]
{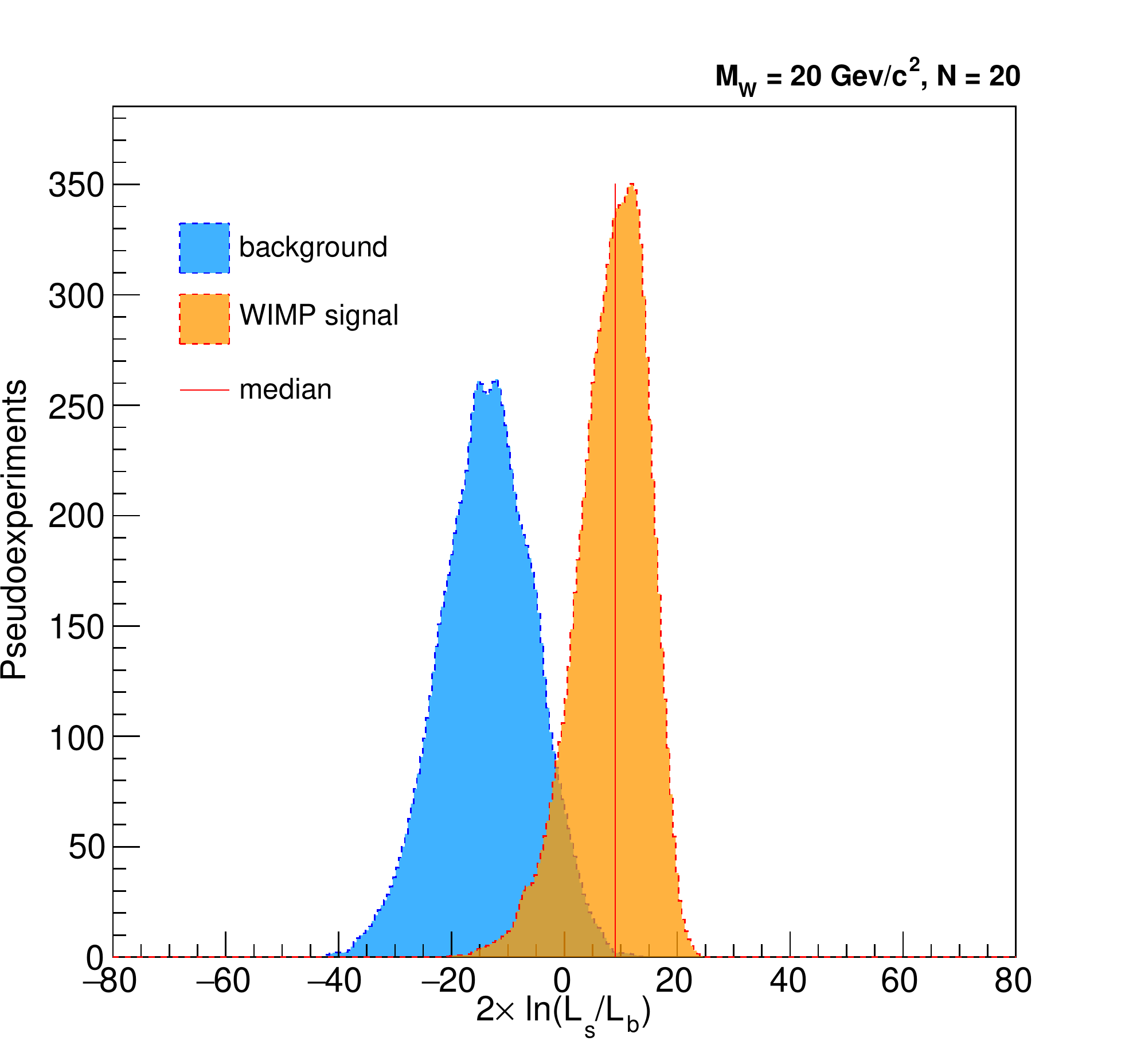}
\end{center}
\caption{\label{fig:test_m20}  Distribution of the test statistics $q= 2 \mbox{ln}( \mathscr{L}_s/ \mathscr{L}_b)$ for the M$_W$=20 GeV/c$^2$ WIMP hypothesis tested against the background only hypothesis, assuming 20 observed events. The expectation for
the background is represented by the blue histogram on the left and for the signal hypothesis by the orange histogram on the right. The red line indicates the median $q$ value for the WIMP signal.}
\end{figure}

Figure \ref{fig:test_m20} shows the distribution of the test statistics 
for the signal and background hypotheses, assuming 20 observed events and a WIMP mass of 20 GeV/c$^2$. 

The expected separation between the two hypotheses ($p$-value) is estimated as the fraction of cases in the background hypothesis ($H_0$) where the test statistic $q$ is above the median value for the WIMP signal ($H_1$). The $p$-value amounts to $1.27\times10^{-3}$, corresponding to a 3$\sigma$ CL in the rejection of the null hypothesis.

\begin{figure}
\centering
\subfigure{\includegraphics[width=7cm]{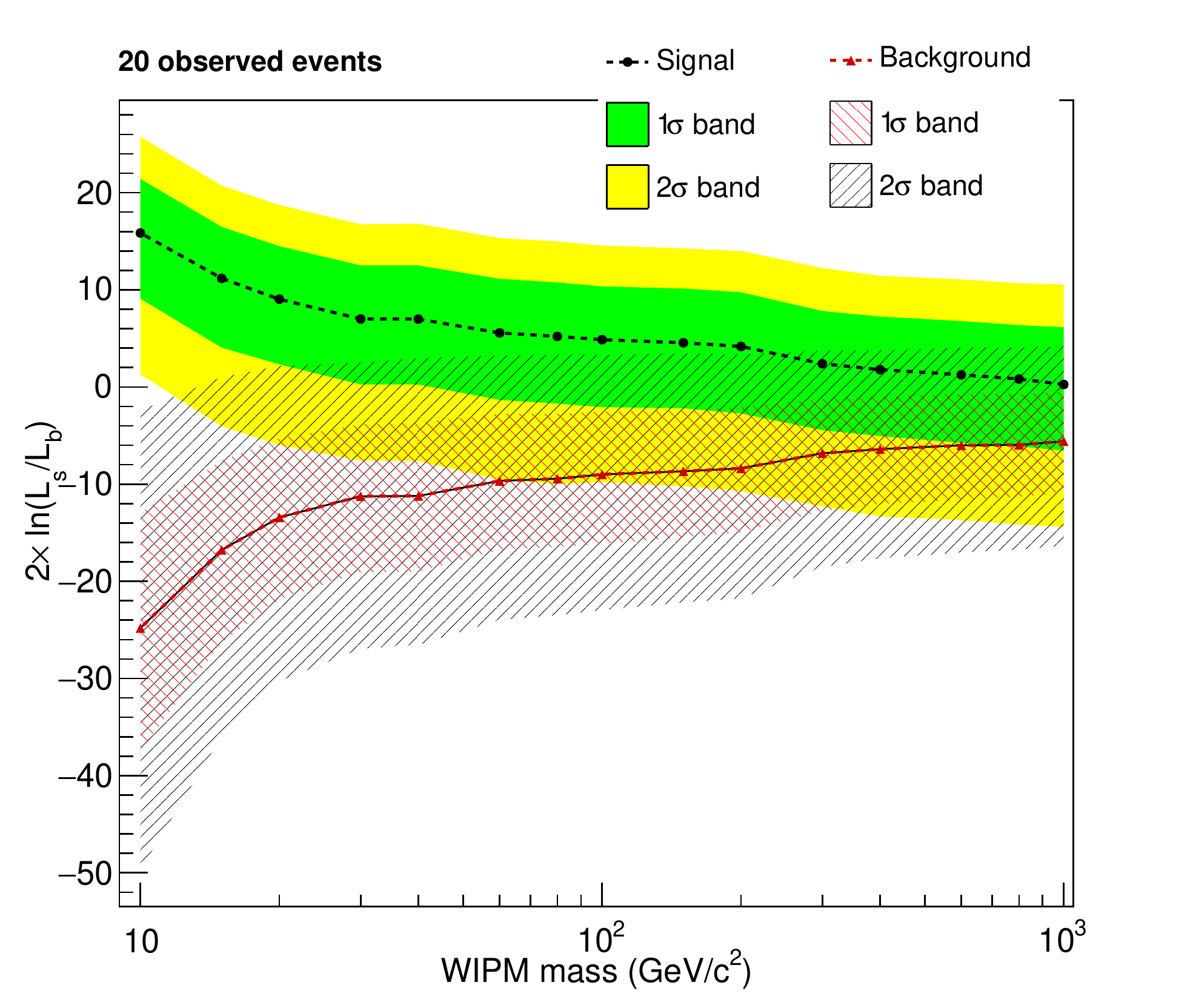}}
\subfigure{\includegraphics[width=7cm]{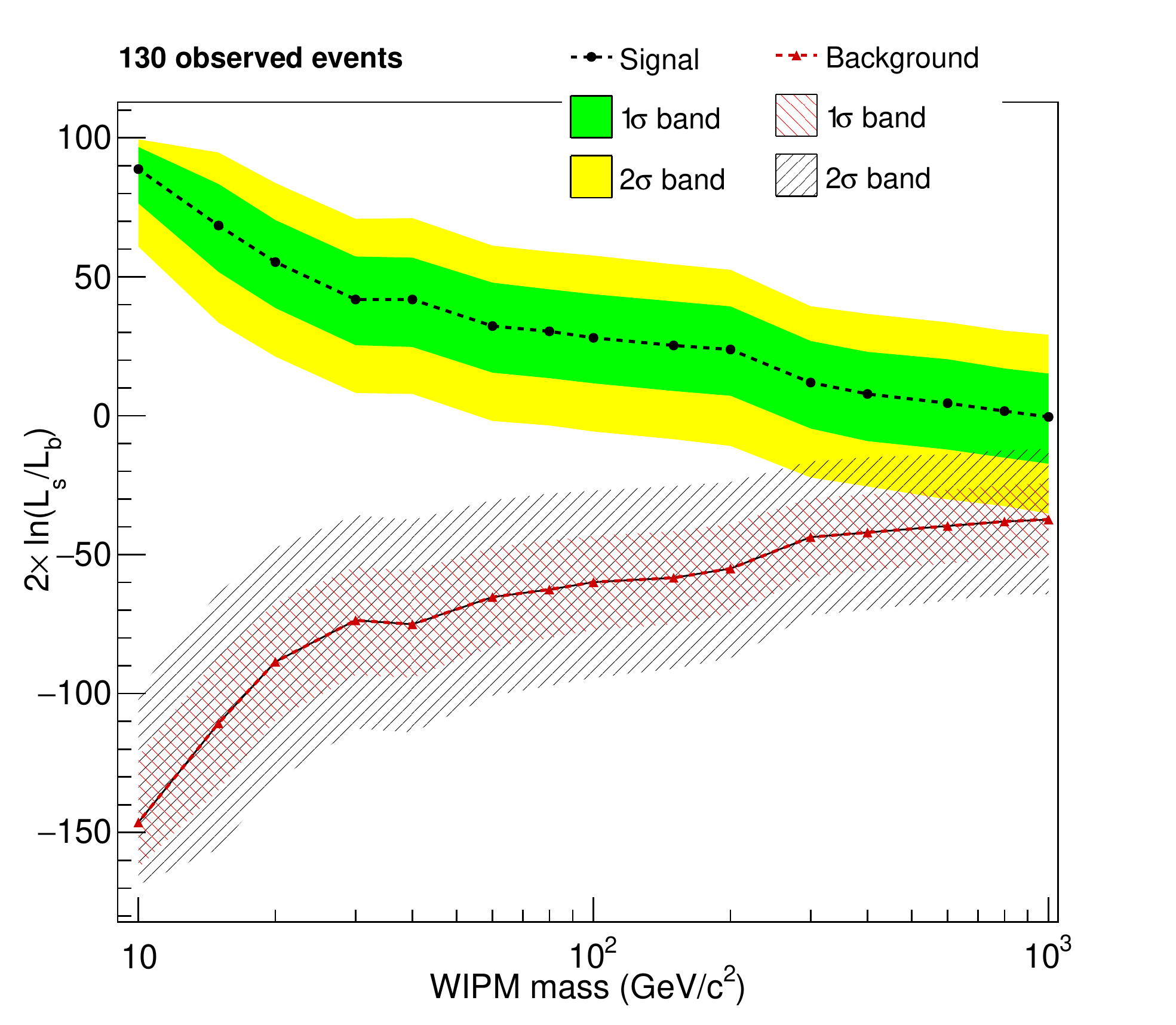}}
\caption{\label{fig:scan}   Test statistic $q$ as a function of the WIMP mass for the signal hypothesis tested against the background hypothesis, assuming 20 (top) or 130 (bottom) observed events.}
\end{figure}

The expected separation for different values of WIMP masses, ranging from 10 to 1000 GeV/c$^2$, is summarized in Figure \ref{fig:scan} assuming 20 (top panel) and 130 (bottom panel) observed events.
The median expectation for the Dark Matter signal is represented by the black dots with the green (68\% CL) and yellow (95\% CL) solid color regions and for the background hypotheses by the red triangles with the red (68\% CL) and black (95\% CL) hatched regions. The observation of 20 (130) events allows to prove that the data are not compatible with the expected background with a 3$\sigma$ CL for WIMP masses below 20 (1000) GeV/c$^2$. 

This highlights the fact that a directional detection approach could lead to the discovery of DM and to the confirmation of its galactic origin with a relatively small number of observed event.

\section{\label{sec:perspectives} NEWSdm towards ``neutrino floor''}

In the previous sections it has been shown how the directional information can be used to discriminate DM signal from isotropic background for a realistic NEWSdm detector with 100 kg $\times$ year exposure and 100 nm threshold.

The discrimination based on the measurement of the recoil direction offers the unique possibility to search for a DM signal beyond the ``neutrino floor'', where neutrinos interacting coherently with atomic nuclei would produce recoils  that cannot be distinguished from DM interactions. 

\begin{figure}[h]
\centering
\begin{center}
\includegraphics[width=7cm]
{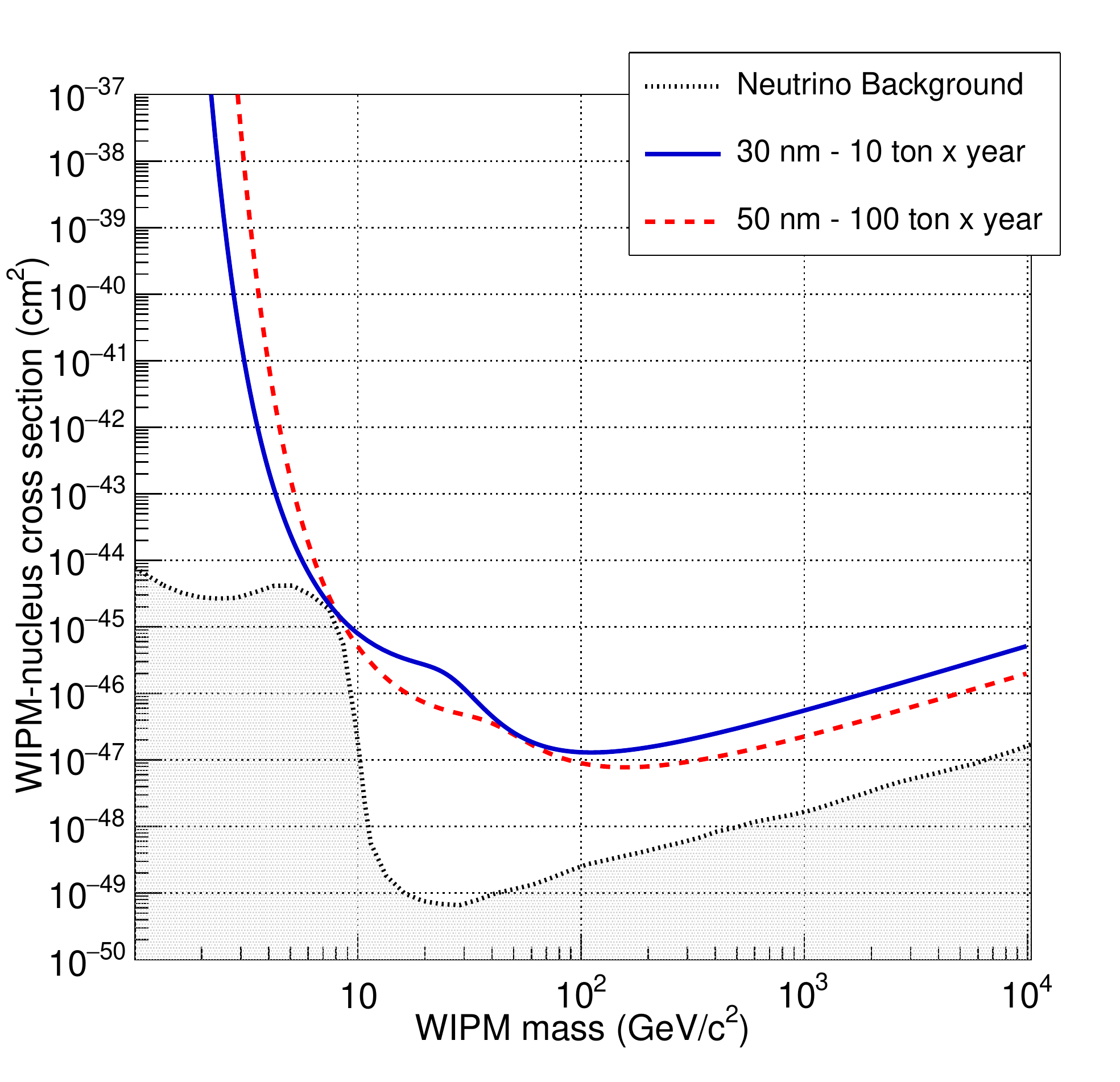}
\end{center}
\caption{\label{fig:neutrino_bkg}  Exclusion limits for the NEWSdm detector with 10 ton $\times$ year exposure and 30 nm threshold (solid blue curve) and with  100 ton $\times$ year exposure and 50 nm threshold (dashed red curve). Zero background is assumed. The gray dotted curve represents the neutrino bound, as evaluated in \cite{Billard}.}
\end{figure}

The exploration of the neutrino background region requires both the construction of a larger mass detector and the reduction of the track length threshold. 
In Figure \ref{fig:neutrino_bkg} the NEWSdm exclusion limit in the zero background hypothesis is compared with the curve representing the neutrino bound for a Xe/Ge target, as evaluated in \cite{Billard}. The neutrino limit is reached with a 10 (100) ton $\times$ year exposure if a 30 (50) nm threshold is assumed.

Thanks to a new technology based on the use of polarized light with optical microscopes, a 10 nm
accuracy has been already achieved on both $X$ and $Y$ coordinates while keeping the high scanning speed performances of fully automated optical microscopes \cite{Aleksandrov}. Therefore, it is certainly possible to lower the minimum detectable track length beyond 100 nm, provided that the grain size is small enough to build a track within that range. 
Recently Ultra-Nano Imaging Tracker (U-NIT) with a grain size of about 20 nm have been developed, paving the way for a reduction of the detector threshold.

\section{\label{sec:concl}Conclusions}
The use of fine-grained nuclear emulsions both as target and nanometric tracking device for directional DM searches offers a unique opportunity for a high-significance discovery of galactic DM. Indeed, directional detection is the only way to extend the DM direct detection beyond the ``neutrino floor''. 

In this paper we have evaluated the discovery potential of the NEWSdm detector in the search for WIMPs originated in the galactic halo, assuming a 100 nm threshold in the track length and the 2D angle reconstruction without head-tail sense recognition. A detailed simulation was performed to take into account the straggling effect of nuclear recoils in the target material. 

As a result of the continuous improvements in the field of nuclear emulsion technologies, relevant progresses in the reduction of the track length threshold are foreseen in the near future, thus allowing to significantly enhance the explored region in the DM parameter space towards and possibly beyond the ``neutrino floor''.


\begin{thebibliography}{9}

\bibitem{Battat}
J.B.R.~Battat et al., Phys.~Rept.~\textbf{C662}, 1 (2016)
\bibitem{Tovey}
D.R.~Toveyet al., Phys.~Lett.~\textbf{B488}, 17 (2000)
\bibitem{Natsume}
M.~Natsume et al., Nucl.~Instrum.~Meth.~\textbf{A575}, 439 (2007)
\bibitem{Naka}
T.~Naka et al., Nucl. Instrum.~Meth.~\textbf{A718}, 519 (2013)
\bibitem{Acquafredda}
R.~Acquafredda et al.~(OPERA Collaboration), JINST \textbf{4}, P04018 (2009)
\bibitem{Aleksandrov}
A.~Aleksandrov et al.~(NEWSdm Collaboration), LNGS-LOI 48/15  (2016), arXiv:1604.04199 [astro-ph.IM]
\bibitem{Ullio}
P.~Ullio and M.~Kamionkowski, JHEP  \textbf{03}, 049 (2001)
\bibitem{Morgan}
B.~Morgan, A.M.~Green and N.J.C.~Spooner, Phys.~Rev.~\textbf{D71}, 103507 (2005)
\bibitem{Aleksandrov2}
A.~Aleksandrov et al., Astropart.~Phys.~\textbf{80}, 16 (2016)
\bibitem{Ziegler}
J.F.~Ziegler, M.~Ziegler and J.~Biersack,  Nucl.~Instrum.~Meth.~\textbf{B268}, 1818 (2010)
\bibitem{Cowan}
G.~Cowan, K.~Cranmer, E.~Gross and O.~Vitells,  Eur.~Phys.~J.~\textbf{C71}, 1554 (2011), [Erratum: Eur.~Phys.~J.~\textbf{C73}, 2501 (2013)]
\bibitem{Moneta}
L.~Moneta et al.,  PoS \textbf{ACAT2010}, 057 (2010)
\bibitem{Verkerke}
W.~Verkerke, CERN-2008-001, 169 (2007)
\bibitem{Brun}
R.~Brun and F.~Rademakers, Nucl.~Instrum-~Meth.~\textbf{A389}, 81 (1997)
\bibitem{Read}
A.L.~Read in \emph{Workshop on confidential limits, CERN, Geneva, Switzerland, 17-18 Jan 2000: Proceedings}, p.81
\bibitem{Billard}
J.~Billard, E.~Figueroa-Feliciano and L.~Strigari, Phys.~Rev.~D \textbf{89}, 023524 (2014)

\end{thebibliography}
\end{document}